\documentclass[12pt]{article}
\usepackage{amssymb}
\makeatletter

\newcommand{\be}{\begin{equation}}
\newcommand{\ee}{\end{equation}}
\newcommand{\bea}{\begin{eqnarray}}
\newcommand{\eea}{\end{eqnarray}}

\def\ie{{\it i.e.}}

\def\thebibliography#1{\paragraph{References\@mkboth
 {REFERENCES}{REFERENCES}}\list
 {[\arabic{enumi}]}{\settowidth\labelwidth{[#1]}\leftmargin\labelwidth
 \advance\leftmargin\labelsep \itemsep=0pt
 \usecounter{enumi}}
 \def\newblock{\hskip .11em plus .33em minus .07em}
 \sloppy\clubpenalty4000\widowpenalty4000
 \sfcode`\.=1000\relax\small}
\makeatother

\input amssym.def       
\input amssym.tex       

\def\H{{\cal H}}
\def\Cop{{\mathbb C}}
\def\Zop{{\mathbb Z}}

\def\Nop{{\mathbb N}}
\def\R{{\cal R}}

\begin{document}

\thispagestyle{empty}
\def\thefootnote{\fnsymbol{footnote}}
\vskip 4.5em
\begin{center}\LARGE
  {\bf Logarithmic torus amplitudes}
\end{center}\vskip 2em
\begin{center}
  Michael Flohr\footnote{Email: {\tt flohr@th.physik.uni-bonn.de}}  \\
{\it Physikalisches Institut, University of Bonn} \\
{\it Nussallee 12, D-53115 Bonn, Germany} \\ \vspace*{0.3cm}
and \\ \vspace*{0.3cm}
Matthias R. Gaberdiel\footnote{Email: {\tt
    gaberdiel@itp.phys.ethz.ch}} \\
{\it Institut f\"ur Theoretische Physik, ETH Z\"urich} \\
{\it ETH-H\"onggerberg, 8093 Z\"urich, Switzerland}
\end{center}
\vskip 1em
\begin{center}
  \today
\end{center}
\vskip 1em
\begin{abstract}
For the example of the logarithmic triplet theory at $c=-2$ the 
chiral vacuum torus amplitudes are analysed. It is found that the
space of these torus amplitudes is spanned by the characters of the
irreducible representations, as well as a function that can be
associated to the logarithmic extension of the vacuum
representation. A few implications and generalisations of this result
are discussed. 
\end{abstract}

\setcounter{footnote}{0}
\def\thefootnote{\arabic{footnote}}

\paragraph{1. Introduction.}

During the last twenty years much has been understood about the
structure of rational conformal field theories. 
Rational conformal field theories are characterised by the property
that they have only finitely many irreducible highest weight 
representations of the chiral algebra (or vertex operator algebra),
and that every highest weight representation is completely
decomposable into irreducible representations. The structure of these
theories is well understood: in particular, the characters of the
irreducible representations transform into one another under modular
transformations \cite{Zhu} (see also \cite{Nahm91}), and the modular
$S$-matrix determines the fusion rules via the Verlinde formula
\cite{Verlinde}. (A general proof for this has only recently been
given in \cite{Huang2005}.)  

On the other hand, it is clear that rational conformal field theories
are rather special, and it is therefore important to understand the
structure of more general classes of conformal field theories. One
such class are the (rational) logarithmic theories that possess only
finitely many indecomposable representations,  
but for which not all highest weight representations are completely 
decomposable. The name `logarithmic' comes from the fact that their
chiral correlation functions typically have logarithmic branch
cuts. The first example of a (non-rational) logarithmic conformal
field theory was found in \cite{Gurarie:1993xq} (see also
\cite{RSal92}), and the first rational example (that shall also
concern us in this paper) was constructed in  \cite{GKau96b}; for some
recent reviews see
\cite{Flohr:2001zs,Gaberdiel:2001tr,Kawai:2002fu}. From a physics 
point of view, logarithmic conformal field theories appear naturally
in various models of statistical physics, for example in 
the theory of (multi)critical polymers \cite{Sal92,Flohr95,Kau95},
percolation \cite{Watts96,Flohr:2005ai}, two-dimensional turbulence
\cite{RR95,Flohr96b,RR96}, the quantum Hall effect
\cite{Gurarie:1997dw} and various critical (disordered) models
\cite{CKT95,MSer96,CTT98,Gurarie:1999yx,Ruelle:2002jy,Piroux:2004vd,%
Moghimi-Araghi:2004wg,deGier:2003dg}. There have also 
been applications in Seiberg-Witten models \cite{Flohr:1998ew} and in     
string theory, in particular in the context of 
D-brane recoil  \cite{KMa95,PT,KMW,Lambert:2003zr}, and in pp-wave
backgrounds  \cite{Bakas:2002qh}. Logarithmic vertex operator algebras
have finally attracted some attention recently in mathematics 
\cite{Milas:2001bb,Kleban:2002pf,Miyamoto,Huang:2002mx,Huang:2003za}.  
Most examples that have been studied concern the $c=-2$ model (that
shall also mainly concern us here), but logarithmic conformal field
theories have also arisen in other contexts, see for example
\cite{Gaberdiel:2001ny,Lesage:2002ch,Rasmussen:2005hj,Fjelstad:2002ei}. 
\smallskip

As we have mentioned above, the characters of the irreducible
representations of a rational conformal field theory close under the
action of the modular group. This can be proven by showing that they
span the space of (chiral) vacuum torus amplitudes which is modular
invariant on general grounds \cite{Zhu}. On the other hand, for
logarithmic conformal field theories it has been known for some time
that the characters of the 
irreducible representations do not, by themselves, form a
representation of the modular group \cite{Flohr95,Flohr96}. However, 
even for logarithmic theories the vacuum torus amplitudes should still
be closed under the action of the modular group \cite{Miyamoto}. In
order to see explicitly how this fits together, we study in this paper
the space of vacuum torus amplitudes for the example of the triplet
theory at $c=-2$ \cite{Kausch91}. We explain how to derive the 
modular differential equation that characterises  these 
amplitudes. (In the case of rational conformal field 
theories, such differential equations were first considered in 
\cite{MMS}.) As we shall see, the characters of the irreducible 
representation only account for a subspace of codimension one.
Furthermore, we show that the remaining solution of the differential
equation can be taken to agree with the `logarithmic  character' that
can be formally associated to the indecomposable extension of the
vacuum  representation \cite{Flohr96}; this clarifies its
interpretation as a genuine vacuum torus amplitude (despite the fact
that it is not actually a character). We also observe that this
association of a vacuum torus amplitude to a logarithmic
representation is not canonical. In particular, the indecomposable
highest weight representations therefore do not give rise to a 
canonical basis for the space of these torus amplitudes. This explains
why Verlinde's formula (that presupposes such a basis) cannot describe
the fusion rules of the triplet theory correctly \cite{GKau96b}.  
 
The modular properties of a logarithmic conformal field theory have
played an important role in various applications of logarithmic
conformal field theory, in particular in the analysis of the boundary
theory (for some work in this direction see 
\cite{Kogan:2000fa,Kawai:2001ur,Bredthauer:2002ct,Pearce:2002an,%
deGier:2005fx}) and the fusion rules \cite{Flohr96,Feigin:2005zx}. 
\smallskip

The paper is organised as follows. In section~2 we review briefly the
main results of Zhu \cite{Zhu} that were generalised to the
logarithmic case in \cite{Miyamoto}. In section~3 we recall the main
properties of the $c=-2$ triplet theory. Putting these results
together we derive, in section~4, the modular differential equation
that characterises the vacuum torus amplitudes. The complete space of
solutions is constructed in section~5. In section~6 we explain how the
analysis of the modular differential equation can be generalised to
arbitrary rational logarithmic conformal field theories. Finally, we
sketch in section~7 how the analysis works for the other triplet
theories, giving explicit details for the $c=-7$ example.

\paragraph{2. Zhu's argument.}

In the following we shall consider conformal field theories (or vertex
operator algebras) that satisfy the $C_2$ condition, but we shall not
assume that they define {\it rational} conformal field theories. As is
common in the mathematical literature, we call a conformal field
theory rational if (i) it possesses only finitely many irreducible
highest weight representations, each of which has finite-dimensional
$L_0$ eigenspaces; and (ii) every highest weight representation can be
decomposed into a direct sum of irreducible highest weight
representations. The $C_2$ condition states that the quotient space
$\H_0 / C_2(\H_0)$ is finite dimensional, where $\H_0$ is the vacuum
representation of the conformal field theory and $C_2(\H_0)$ is the
space spanned by the states  
\be\label{span}
V_{-h(\psi)-1}(\psi)\, \chi \,, \qquad \hbox{for $\psi,\chi\in\H_0$.}
\ee
The $C_2$ condition implies that Zhu's algebra $A(\H_0)$ is finite
dimensional, and therefore that the conformal field theory has only
finitely many irreducible highest weight representations (see also
\cite{gabgod} for an introduction to these matters). However, it does
not imply that the theory is rational in the above sense. Indeed, the
example we shall mainly consider in this paper, the triplet algebra at
$c=-2$ \cite{Kausch91}, satisfies the $C_2$ condition \cite{Nils},
yet is not rational since it possesses indecomposable
representations \cite{GKau96b}.
It is natural to conjecture\footnote{A related conjecture was
originally made by one of us (MRG) in collaboration with Peter Goddard
--- see  \cite{Gaberdiel:2001tr}.} that rational logarithmic conformal
field theories are characterised by the condition that they are
$C_2$-cofinite, but that Zhu's algebra is not semisimple. The results
of this paper are certainly in agreement with this idea. 

Let us briefly summarise the key results of Zhu \cite{Zhu} that were 
extended by Miyamoto \cite{Miyamoto} to theories that satisfy the
$C_2$ condition but are not rational in the above sense. 
If the conformal field theory satisfies the $C_2$ condition, then 
every highest weight representation gives rise to a torus amplitude;
in particular, the vacuum torus amplitude is just given by the usual
character  
\be\label{char}
\chi_{\H_j}(\tau) = 
\hbox{Tr}_{\H_j} \left( q^{L_0-{c\over 24}} \right) \,, \qquad
q= e^{2\pi i \tau} \,,
\ee
which converges absolutely for $0 < |q| < 1$. Furthermore, the space
of torus amplitudes is finite dimensional, and it carries a
representation of SL$(2,\Zop)$ \cite{Zhu,Miyamoto}. As is explained in
\cite{Zhu,Miyamoto}, if the conformal field theory satisfies the $C_2$
condition then there exists a positive integer $s$ so that every
vacuum torus amplitude $T(q)$ satisfies
\be\label{diff}
\left[ \left(q {{\rm d}\over {\rm d}q}\right)^s + \sum_{r=0}^{s-1} h_r(q) 
 \left(q {{\rm d}\over {\rm d}q}\right)^r \right] T(q) = 0 \,.
\ee
Here the $h_r(q)$ are polynomials in the Eisenstein series 
$E_2(q)$, $E_4(q)$ and $E_6(q)$; we choose the convention that the 
Eisenstein series are defined by  
\begin{eqnarray}
E_k(q) &=& 1 - \frac{2\, k}{B_k}
\sum_{n=1}^{\infty}\sigma_{k-1}(n)\, q^n\,,\\
\sigma_k(n) &=& \sum_{d|n}d^k\,,
\end{eqnarray}
where $B_k$ is the $k$-th Bernoulli number. Thus, the $q$-expansion of
the Eisenstein series reads $E_2 = 1 - 24q-72q^2-96q^3-\cdots$,  
$E_4 = 1 + 240q + 2160q^2 + 6720q^3+\cdots$, and 
$E_6 = 1 - 504q - 16632q^2 -122976q^3-\cdots$ in our normalisation. 

For the following it is important (see Lemma 5.3.2 of \cite{Zhu}) 
that the functions $h_r$ have the property that 
\be\label{rel}
\left(L-{c\over 24}\right)^s + \sum_{r=0}^{s-1} h_r(0) 
\left(L-{c\over 24}\right)^r = 0 
\ee
in Zhu's algebra $A(\H_0)$. This reflects the fact that for
$q\rightarrow 0$, only the highest weight states contribute to the
vacuum torus amplitudes, and that they must therefore satisfy the
constraints of Zhu's algebra. As we shall argue below, the
differential equation (\ref{diff}) can be identified with the modular
differential equation that was first considered in \cite{MMS}.      

If the conformal field theory is in addition rational in the above
sense Zhu showed that the space of torus amplitudes is spanned
by the characters of the irreducible representations. However, as
already pointed out in \cite{Miyamoto}, this is not longer the case if
the theory is not rational. Indeed, we shall see this very explicitly
for the case of the triplet algebra in the following.

\paragraph{3. The triplet theory.}

Let us briefly recall some of the properties of the triplet theory
\cite{Kausch91,Flohr95,Kau95,GKau96b}. The chiral algebra for this
conformal field theory is generated by the Virasoro modes $L_n$, and
the modes of a triplet of weight 3 fields $W^a_n$. The commutation
relations are   
\begin{eqnarray}
  {}[ L_m, L_n ] &=& (m-n)L_{m+n} - \frac16 m(m^2-1) \delta_{m+n}\,, 
  \nonumber \\
  {}[ L_m, W^a_n ] &=& (2m-n) W^a_{m+n}\,, 
  \nonumber \\
  {}[ W^a_m, W^b_n ] &=& g^{ab} \biggl( 
  2(m-n) \Lambda_{m+n} 
  +\frac{1}{20} (m-n)(2m^2+2n^2-mn-8) L_{m+n} 
  \nonumber\\&&\qquad
  -\frac{1}{120} m(m^2-1)(m^2-4)\delta_{m+n}
  \biggr) 
  \nonumber \\&&
  + f^{ab}_c \left( \frac{5}{14} (2m^2+2n^2-3mn-4)W^c_{m+n} 
    + \frac{12}{5} V^c_{m+n} \right)\,, \nonumber
  \end{eqnarray}
where $\Lambda = \mathopen:L^2\mathclose: - 3/10\, \partial^2L$ and 
$V^a = \mathopen:LW^a\mathclose: - 3/14\, \partial^2W^a$ are
quasiprimary normal ordered fields. $g^{ab}$ and $f^{ab}_c$ are the
metric and structure constants of $\mathfrak{su}(2)$. 
In an orthonormal basis we
have $g^{ab} = \delta^{ab}, f^{ab}_c = i\epsilon^{abc}$. 

The triplet algebra (at $c=-2$) is only associative, because certain
states in the vacuum representation (which would generically violate
associativity) are null. The relevant null vectors are
\begin{eqnarray} 
  N^a &=& 
  \left(2 L_{-3} W^a_{-3} -\frac43 L_{-2} W^a_{-4} + W^a_{-6}\right)
  \Omega\,, 
\label{eq:nulllw}
\\
  N^{ab} &=&
  W^a_{-3} W^b_{-3} \Omega - 
  g^{ab} \left( \frac89 L_{-2}^3 + \frac{19}{36} L_{-3}^2 +
  \frac{14}{9} L_{-4}L_{-2} - \frac{16}{9} L_{-6} \right)\Omega 
  \nonumber\\*&&
  - f^{ab}_c\left( -2 L_{-2} W^c_{-4} + \frac54 W^c_{-6}
  \right) \Omega \,.
\label{eq:nullww}
\end{eqnarray}
We shall only be interested in representations which respect these
relations, and for which the spectrum of $L_0$ is bounded from below.
Evaluating the constraint coming from (\ref{eq:nullww}), we find (see
\cite{GKau96b} for more details) 
\begin{equation}\label{3}
  \left(W^a_0 W^b_0 - g^{ab} \frac19 L_0^2 (8L_0 + 1) - 
  f^{ab}_c \frac15 (6L_0-1) W^c_0 \right) \psi = 0 \,,
\label{eq:wwzero}
\end{equation}
where $\psi$ is any highest weight state, while the relation coming
from the zero mode of (\ref{eq:nulllw}) is satisfied identically. 
Furthermore, the constraint from $W^a_1 N^{bc}_{-1}$, together with
(\ref{eq:wwzero}) implies that $W^a_0 (8 L_0 - 3) (L_0 -1) \psi = 0$. 
Multiplying with $W_0^a$ and using (\ref{eq:wwzero}) again, this
implies that 
\begin{equation}
\label{eq:heigen}
0  = L_0^2 (8 L_0 + 1) (8 L_0 - 3) (L_0 - 1) \psi\,.
\end{equation}
For irreducible representations, $L_0$ has to take a fixed value $h$
on the highest weight states, and (\ref{eq:heigen}) then implies that
$h$ has to be either $h=0, -1/8, 3/8$ or $h=1$. However, it also
follows from (\ref{eq:heigen}) that a logarithmic highest weight 
representation is allowed since we only have to have that $L_0^2=0$
but not necessarily that $L_0=0$. Thus, in particular, a
two-dimensional space of highest weight states with relations
\be\label{3.16}
L_0\, \omega = \Omega \qquad \qquad L_0\, \Omega = 0 \,.
\ee
satisfies (\ref{eq:heigen}). This highest weight space gives rise to
the `logarithmic' (indecomposable) representation $\R_0$ (see
\cite{GKau96b} for more details). [The other indecomposable
representation $\R_1$ of \cite{GKau96b} is not highest weight in the
usual sense.]

It follows from the above analysis (and a similar analysis for the
$W^a$ modes --- see for example \cite{GKau96b}) that the triplet
theory has only finitely many indecomposable highest weight
representations. This suggests that it satisfies the $C_2$ condition,
and this can be confirmed by a computer calculation
\cite{Kaupriv} (see also \cite{Nils}). Indeed, the space 
$\H_0 / C_2(\H_0)$ has dimension 
$11$, and it can be taken to be spanned by the vectors
\begin{eqnarray}
& L_{-2}^s\Omega\,,   & \hbox{where $s=0,1,2,3,4$} \nonumber \\
& L_{-2}^s W^a_{-3}\Omega \,, \qquad  & \hbox{where $s=0,1$ and
$a\in$ adj($\mathfrak{su}(2))$}\,. \label{C2space}
\end{eqnarray}
As was already explained by Zhu \cite{Zhu}, the dimension of this
quotient space gives an upper bound on the dimension of Zhu's algebra,
which is thus at most $11$-dimensional. On the other hand, it also
follows from the analysis of Zhu \cite{Zhu} that each irreducible
representation whose space of ground states has dimension $d$,
contributes $d^2$ states to Zhu's algebra. For the triplet algebra,
the irreducible representations with highest weights $h=-1/8$ and
$h=0$ are singlet representations, while the irreducible
representations with $h=3/8$ and $h=1$ are doublets. These irreducible
representations  therefore account for a
$1^2+1^2+2^2+2^2=10$-dimensional (sub)space of Zhu's 
algebra. Since we have one additional highest weight representation
--- the logarithmic extension of the vacuum
representation ---  we expect that Zhu's algebra is precisely
11-dimensional, and that the remaining state accounts for this
logarithmic extension. We shall see below how this counting is in fact
mirrored by our analysis of the vacuum torus amplitudes.

\paragraph{4. The modular differential equation.}

The above calculation leading to (\ref{eq:heigen}) implies that in
Zhu's algebra we have the relation 
\be\label{triprel}
L_0^2 (8 L_0 + 1) (8 L_0 - 3) (L_0 - 1) = 0 \,,
\ee
where $L_0$ denotes the operator corresponding to the stress energy
tensor, and the product is to be understood as the product in Zhu's
algebra (see for example \cite{gabgod} for an explanation of this
construction). In fact, such a relation had to hold in Zhu's algebra,
given the structure of the homogeneous quotient space $\H_0/C_2(\H_0)$
in (\ref{C2space}): it follows from (\ref{C2space}) that 
$L_{-2}^5=0$ in $\H_0/C_2(\H_0)$. By the usual argument (see for
example \cite{Zhu}), one can then show that 
\be\label{q1}
L_0^5 + \Bigl(\hbox{terms of conformal weight $< 10$}\Bigr) =  0 
\ee
in Zhu's algebra. The terms of lower conformal weight can again be
expressed in terms of the basis vectors of (\ref{C2space}), as well as
elements in $C_2(\H_0)$. Since all vectors that appear in (\ref{q1}) 
are $\mathfrak{su}(2)$ singlets, only the basis vectors in the first
line of (\ref{C2space}) contribute. Continuing this argument
recursively, one then deduces that there is a fifth order polynomial
relation involving only $L_0$ in Zhu's algebra, \ie\ a relation of the
form (\ref{triprel}).  

By the same token, it then also follows that the differential
equation (\ref{diff}) that characterises the  vacuum torus amplitudes
for the triplet theory is (at most of) fifth 
order. Furthermore, (\ref{rel}) must actually reduce to
(\ref{triprel}), and thus the differential equation is precisely fifth
order. Since the space of vacuum torus amplitudes is invariant 
under the action of the modular group SL$(2,\Zop)$ (see section~2),
the differential equation must be modular invariant as well. The most
general modular invariant differential equation of degree five is  
\be\label{ansatz}
\left[ D^5 + \sum_{r=0}^{4} f_r(q)\, D^r \right] T(q) = 0 \,,
\ee
where each $f_r(q)$ is a polynomial in $E_4(q)$ and $E_6(q)$ of
modular weight $10-2r$, and 
\begin{equation}
D^i = {\it cod}_{(2i)}\cdots{\it cod}_{(2)} {\it cod}_{(0)}\,,
\end{equation}
with $cod_{s}$ being the modular covariant derivative on weight $s$ 
modular functions
\begin{equation}
{\it cod}_{(s)} = q {\partial \over \partial q}
                - \frac{1}{12} (s-2) E_2(q)\,,
\end{equation}
which increments the weight of a modular form by 2. Here $E_2$ is the
second Eisenstein series, and ${\it cod}_{(0)}f=f$. For the case of
rational conformal field theories, this differential equation was
first considered in \cite{MMS} (see also \cite{Eholzer,ES} for further
developments). It is often called the {\it modular differential
equation}.   

The first few of the $D^i$ read to first order in $q$, \ie\ where 
$E_2(q)$  is only taken as $1-24q+{\cal O}(q^2)$, and with the
notation  $D_q=q\frac{\partial}{\partial q}$, simply
\begin{eqnarray*}
  D^0 &=& 1\,,\\
  D^1 &=& D_q^{}\,,\\
  D^2 &=& D_q^2 - \frac16 D_q + q\,4D_q\,,\\
  D^3 &=& D_q^3 - \frac12 D_q^2 + \frac{1}{18}D_q
       + q\,\left(12D_q^2+\frac43 D_q\right)\,,\\
  D^4 &=& D_q^4 - D_q^3 + \frac{11}{36}D_q^2 - \frac{1}{36}D_q
       + q\,\left(24D_q^3 +\frac43D_q^2+\frac43D_q\right)\,,\\
  D^5 &=& D_q^5 - \frac53 D_q^4 + \frac{35}{36}D_q^3 - \frac{25}{108}D_q^2 
                + \frac{1}{54}D_q
       + q\,\left(40D_q^4-\frac{20}{3}D_q^3+\frac{20}{3}D_q^2\right)\,,
\end{eqnarray*}
where all expressions are up to ${\cal O}(q^2)$. Of course, $D^0$ and
$D^1$ are exact to all orders. 

The most general ansatz for the differential equation (\ref{ansatz})
is therefore 
\be\label{ansatz1}
\sum_{k=0}^5\sum_{{r,s\atop4r+6s=10-2k}} 
    a_{r,s}(E_4)^r(E_6)^s\left(
        \prod_{m=0}^{k}{\it cod}_{(2m)}\right) T(q)=0\,.
\ee
This differential equation must be satisfied by the characters of the
irreducible highest weight representations of the triplet algebra. As
we have explained before, there are four irreducible highest weight 
representations with conformal weights $h=0, -1/8, 3/8$ and $h=1$, and
their corresponding characters are known \cite{Flohr95,Kau95,Flohr96}. 
In terms of the functions 
\be\label{etadef}
\eta(q)=q^{1/24}\prod_{n=1}^{\infty} (1-q^n)\,, 
\ee
\be\label{thetalk}
\theta_{\lambda,k}(q)= \sum_{n\in\Zop} q^{{(2kn+\lambda)^2\over4k}}
\,,
\ee
as well as 
\be\label{dtheta}
(\partial\theta)_{\lambda,k}(q)=
\sum_{n\in\Zop} (2kn+\lambda) q^{{(2kn+\lambda)^2\over 4k}} \,,
\ee
they are given as 
\begin{eqnarray}\label{chi:h-18}
\chi_{-{1\over 8}}(q) &=& \theta_{0,2}(q)/\eta(q)\,,\label{car1}\\
\chi_0(q) &=& (\theta_{1,2}(q)+(\partial\theta)_{1,2}(q))/\eta(q)\,,
\label{car2}\\  \label{chi:h38}
\chi_{{3\over 8}}(q) &=& \theta_{2,2}(q)/\eta(q)\,,\label{car3}\\
\chi_1(q) &=& (\theta_{1,2}(q)-(\partial\theta)_{1,2}(q))/\eta(q)\,.
\label{car4}
\end{eqnarray}
Putting these pieces of information together we find that (up to an
overall normalisation constant) (\ref{ansatz1}) is uniquely determined
to be the differential equation
\begin{eqnarray}
0 & =  & \left[\frac{143}{995328}E_4(q) E_6(q) +
   \frac{121}{82944}(E_4(q) )^2{\it cod}_{(2)} +
    \frac{65}{2304}E_6(q) {\it cod}_{(4)}{\it cod}_{(2)}
    \right.\nonumber\\
& &
\left.\mbox{}-\frac{163}{576}E_4(q) {\it cod}_{(6)}{\it cod}_{(4)}
                        {\it cod}_{(2)} +
        {\it cod}_{(10)}{\it cod}_{(8)}{\it cod}_{(6)}{\it cod}_{(4)}
                        {\it cod}_{(2)}\right]T(q)\,.
                        \nonumber
\end{eqnarray}
It is instructive to look at the leading order of the above equation.
If we expand the Eisenstein series $E_n = 1 + g_{n,1}q + {\cal O}(q^2)$ 
with $g_{n,1}$ given by $g_{2,1}=-24,g_{4,1}=240,g_{6,1}=-504$, we
obtain 
\begin{eqnarray*}
  0 &=&
  \left(D_q^5 - \frac53 D_q^4 + \frac{397}{576} D_q^3 -
  \frac{427}{6912}D_q^2    -\frac{37}{82944}D_q +
  \frac{143}{995328}\right)T(q) \\ 
    &+& q\left(40 D_q^4 - \frac{895}{12}D_q^3  + \frac{2209}{96}D_q^2
    - \frac{209}{216}D_q - \frac{1573}{41472}\right)T(q)
    + {\cal O}(q^2)\,.
\end{eqnarray*}
The zero-order term in $q$ can be factorised as
\be\label{lowest}
  \frac{1}{995328}(24D_q-11)(12D_q-13)(24D_q+1)(12D_q-1)^2\,.
\ee
Recalling that $D_q$ has to be replaced by 
$L_0-{c\over 24}=L_0+{1\over 12}$ in order to relate
(\ref{diff}) to (\ref{rel}), this therefore reduces, as required,
to (\ref{triprel}). If we make the ansatz
\be\label{torusans}
T(q) = q^{h+{1\over 12}} \left( 1 + c_1 q + c_2 q^2 + c_3 q^3 + 
{\cal O}(q^4) \right)\,,
\ee
the above differential equation becomes, up to third order,
\begin{eqnarray*}
  0&=&\frac{q^{h+1/12}}{64}\left[q^0\left(h^2(h-1)(8h+1)
                                (8h-3)\right)\right.\\
   &+& q^1\left(c_1(h+1)^2h(8h+9)(8h+5) 
             + 2h(32h-45)(40h^2-5h-1)\right)\\
   &+& q^2\left(c_2(h+2)^2(h+1)(8h+17)(8h+13) 
             + 2c_1(32h-13)(h+1)(40h^2+75h+34)\right.\\
   & & \left.\phantom{ml}+ 2(3840h^4+2840h^3-17331h^2+706h-442)\right)\\
   &+& q^3\left(c_3(h+3)^2(h+2)(8h+25)(8h+21) 
             + 2c_2(h+2)(32h+19)(40h^2+155h+149)\right.\\
   & & \left.\phantom{ml}+ 2c_1(3840q^4+18200q^3
                       +14229q^2-10076q-10387)\right.\\
   & & \left.\left.\phantom{ml}+ 4(2560h^4+28880h^3
                       -66574h^2-9772h-12281)\right)
       + {\cal O}(q^4)\right]\,.
\end{eqnarray*}

\paragraph{5. Solving the modular differential equation.}

As we have argued above, the modular differential equation is of fifth 
order for the triplet theory, and the space of vacuum torus amplitudes 
is therefore five-dimensional. On the other hand, we have only got
four irreducible representations that give rise, via their characters,
to four vacuum torus amplitudes (that solve the differential 
equation). Let us now analyse how to obtain a fifth, linearly
independent, vacuum torus amplitude. First let us try to find a
solution of the form (\ref{torusans}). Because of the lowest order
equation (\ref{lowest}), this will only give rise to a solution
provided that $h=-{1\over 8}, {3\over 8}, 0$ or $h=1$. For each fixed
$h$, one then finds that there is only one such solution, which
therefore agrees with the corresponding character of the irreducible 
representation (\ie\ with (\ref{car1}) -- (\ref{car4})). By the way,
this conclusion was not automatic {\it a priori}, since there exist
cases where the modular differential equation has two linearly
independent solutions with the same conformal weight, both of which
are of power series form. The simplest example is provided by the two
$h=0$ characters of the $c=1-24k$ series of rational CFTs, $k\in\Nop$,
with extended symmetry algebra ${\cal W}(2,8k)$. One of these
solutions belongs to the vacuum representation, the other to a second
$h=0$ representation (whose highest weight state has a
non-zero $W_0$ eigenvalue).

The character of any highest weight representation always gives rise
to a torus amplitude as in (\ref{torusans}), and thus we have shown
that the space of vacuum torus amplitudes for the triplet theory 
is not spanned by the characters of the (irreducible) highest weight
representations. This was to be expected, given the analysis of
\cite{Miyamoto}. 

It is not difficult to show that the missing, linearly 
independent solution can be taken to be 
\begin{equation}\label{T5}
T_5(q) = \log(q)(\partial\theta)_{1,2}(q)/\eta(q)\,.
\end{equation}
It is tempting to associate this vacuum torus amplitude with the
logarithmic (indecomposable) highest weight representation $\R_0$ 
whose ground state conformal weight is $h=0$, and this is indeed what
was suggested in \cite{Flohr96}. However, strictly speaking, this
identification is only formal since $T_5(q)$ is not canonically
determined by the above analysis. In particular, we could have equally
replaced $T_5(q)$ by   
$T'_5(q) = T_5(q)+\alpha_0 \chi_0(q) + \alpha_1 \chi_1(q)$ for any
(real)  $\alpha_i$, $i=1,2$. It is  therefore not clear which choice
of the $\alpha_i$ should (formally) describe the character of the
logarithmic representation $\R_0$.  [It is also clear that the
conventional character of ${\cal R}_0$ is in fact just 
\begin{equation}
  \chi_{{\cal R}_0}^{}(q) = \chi_0^{}(q)+\chi_1^{}(q) 
                          = 2\theta_{1,2}(q)/\eta(q)\,,
\end{equation}
and therefore does not account for the additional solution. The same
is also true for the other indecomposable representation 
${\cal R}_1$.] 

One important consequence of this analysis is that the space of torus
amplitudes does not have a canonical basis. This is unlike the case of
a rational conformal field theory where the canonical basis for the
space of vacuum torus amplitudes is given in terms of the characters
of the irreducible representations. This canonical basis plays a
crucial role in the Verlinde formula, where the matrix elements of the
modular $S$-matrix with respect to this basis enters. It is therefore
not surprising that the Verlinde formula does not work for this
logarithmic conformal field theory: as was shown in
\cite{GKau96b}, the fusion rules of the triplet theory cannot be
diagonalised, and thus no Verlinde formula can exist.

Finally, we note that the solution $T_5(q)$ is in fact a torus
amplitude in a slightly different sense. As we have seen above,
$T_5(q)$ is proportional to $\tau\eta^2(q)$ and thus, up to the
Liouville factor, proportional to one of the periods of the torus.
Following an approach of Knizhnik \cite{Kniz}, one can show that this
torus amplitude is precisely one of the two conformal blocks one finds
for the four-point function $\langle\mu\mu\mu\mu\rangle$ of the
$h=-1/8$ field on the plane, provided we express it in terms of $\tau$
instead of the crossing ratio $x$ with the help of the elliptic modulus
$\kappa^2(\tau)=x$, see \cite{Floh04}. Actually, this four-point
function gives the complex plane the geometry of a double covering
with two branch cuts, \ie\ of a torus.

\paragraph{6. A general analysis.}

For any (logarithmic) conformal field theory which satisfies the $C_2$
condition, the mere existence of a finite order differential equation
allows us to derive some relations and bounds for the highest weights.
As argued above, the torus amplitudes of such a theory have to satisfy
an $n$-th order holomorphic modular invariant differential equation of 
the form (\ref{ansatz}), 
\be\label{ansatz2}
\left[D^n + \sum_{r=0}^{n-1}f_r(q)D^r\right]T(q) = 0\,,
\ee
where the $f_r(q)\in\Cop[E_4,E_6]$ are modular functions of weight
$2(n-r)$. These coefficient functions may be expressed in terms of a
set of $n$ linearly independent solutions $T_1(q),\ldots,T_n(q)$ of
the differential equation (\ref{ansatz2}). However, in
contrast to \cite{MMS}, these solutions cannot in general be
identified with the characters of representations. In particular, we
cannot assume that the $T_i(q)$ have a good power series 
expansion in $q$ up to a common fractional power 
$h_i - c/24\ {\rm mod}\ 1$.\footnote{We will in the following always
speak of power series expansions in $q$ with the silent understanding
that a common fractional power is allowed, \ie\ that 
the functions can be expanded as 
$T(q)=q^\alpha\sum_{k=0}^{\infty} a_kq^k$, $\alpha\in\mathbb{Q}$.}
Instead we want to assume that they lie in 
$\Cop(\!(q)\!)[\tau]$, \ie\ that they are power series in $q$
times a polynomial in $\tau\equiv{1\over 2\pi{\rm i}}\log(q)$. This
is certainly the case for the triplet theory discussed before.

With this in mind we can adapt the analysis of \cite{MMS} to this more
general setting. The main difference will be that we shall {\em not\/}
assume in the following that the highest weights are all different,
$h_i\neq h_j$ for $i\neq j$, but only that $T_i(q)\neq T_j(q)$ for
$i\neq j$. Note that the asymptotic behaviour of two functions
$T_i(q)$ and $T_j(q)$ in the limit $q\rightarrow 0$ (or
$\tau\rightarrow+{\rm i}\infty$) is the same whenever
$T_j(q)=p(\tau)T_i(q)$ for a polynomial $p$, provided  
$T_i(q)\sim q^\alpha$ with $\alpha\neq 0$. The case $\alpha=0$ occurs
precisely when $h_i-c/24=0$.  We note that all known logarithmic
conformal field 
theories, except for $c=0$, do not have any logarithmic
representations with $h=c/24$. 
\smallskip

As in \cite{MMS} we now express the coefficients of the
modular differential equation in terms of the Wronskian of a set of
$n$ linearly independent solutions as
\begin{eqnarray}
  f_r(q) &=& (-1)^{n-r}W_r(q)/W_n(q)\,,\\   \label{eq:Wn}
  W_r(q) &=& {\rm det}\left(\begin{array}{rcr}
         T_1(q) & \phantom{m}\ldots\phantom{m} &        T_n(q) \\
      D^1T_1(q) &            \ldots            &     D^1T_n(q) \\
    \vdots\phantom{m}  &                   & \vdots\phantom{m} \\
  D^{r-1}T_1(q) &            \ldots            & D^{r-1}T_n(q) \\
  D^{r+1}T_1(q) &            \ldots            & D^{r+1}T_n(q) \\
    \vdots\phantom{m}  &                   & \vdots\phantom{m} \\
      D^nT_1(q) &            \ldots            &     D^nT_n(q) 
  \end{array}\right)\,.
\end{eqnarray}
The torus amplitudes, considered as functions in $\tau$, are
non-singular in $\mathbb{H}$. As a consequence, the same applies for
the $W_r$. Therefore, the coefficients $f_r$ can have singularities
only at the zeroes of $W_n$.  We will now show that the total number
of zeroes of $W_n$ can be expressed in terms of the number $n$ of 
linearly independent torus amplitudes, the central charge $c$ and the
conformal weights $h_i$ associated to the torus amplitudes
$T_i(q)$. In order to do so, we note that in the 
$\tau\rightarrow+{\rm i} \infty$ limit, the torus amplitudes behave as
$\exp(2\pi{\rm i}(h_i - {c\over 24})\tau)$. With the above caveat 
concerning the case $h=c/24$, this applies to all 
torus amplitudes independently of whether they are pure power series
in $q$, or whether they have a $\tau$-polynomial as additional
factor. This implies that 
$W_n\sim\exp(2\pi{\rm i}(\sum_ih_i - n{c\over 24})\tau)$, which says
that $W_n$ has a pole of order $n{c\over 24}-\sum_ih_i$ at
$\tau={\rm i}\infty$.  Now, $W_n$ involves precisely 
${1\over   2}n(n-1)$ derivatives meaning that it transforms as a
modular form of weight $n(n-1)$. Both facts together allow us to
compute the total number of zeroes of $W_n$, which is  
\be\label{zeroes}
  {1\over 6}\ell\equiv 
  -\sum_{i=1}^n h_i + {1\over 24}nc + {1\over 12}n(n-1)\geq 0\,,\qquad
  \ell\in\mathbb{Z}_+-\{1\}\,.
\ee
This number cannot be negative since $W_n$ must not have a pole in the 
interior of moduli space. We note that (\ref{zeroes}) is always a
multiple of ${1\over 6}$ since $W_n$, as a single valued function in
Teichm\"uller space, may have zeroes at the ramification points
$\exp({1\over 3}\pi{\rm i})$ and $\exp({1\over 2}\pi{\rm i})$ of order
${1\over 3}$ and ${1\over 2}$, respectively. Equation (\ref{zeroes})
provides a simple bound on the sum of the conformal weights. 

For example, for the case of the $c=-2$ triplet theory, we have 
\be
  -\left[(-{1\over 8})+(0)+(0)+({3\over 8})+(1)\right] 
      + {1\over24}(5)(-2) + {1\over 12}(5)(4) = 0\,,
\ee
in agreement with the above analysis.

\paragraph{7. The other triplet theories.} The analysis presented so
far  can in principle be generalised to all members of the $c_{p,1}$
series of triplet models. In practice, however, we have not found it
possible to give uniform explicit expressions.
The pattern which emerges in the treatment of the $c=-2$ case, \ie\
the case $p=2$, however, seems to be of a generic nature. Indeed, all
the $c_{p,1}$ models are $C_2$ cofinite \cite{Nils} and the characters
of their irreducible representations are all known. They close under
modular transformations provided that a certain number of `logarithmic
vacuum torus amplitudes' (the analogues of $T_5(q)$) are added to the
set. In fact, the characters of the irreducible representation,
together with  additional torus amplitudes which we may again
associate to the indecomposable representations, read \cite{Flohr96}  
\bea
  \chi^{}_{0,p}(q)      & = & \frac{1}{\eta(q)}\Theta_{0,p}(q)\,,\\
  \chi^{}_{p,p}(q)      & = & \frac{1}{\eta(q)}\Theta_{p,p}(q)\,,\\
  \chi^+_{\lambda,p}(q) & = & \frac{1}{p\eta(q)}\left[
                            (p-\lambda)\Theta_{\lambda,p}(q)
                          + (\partial\Theta)_{\lambda,p}(q)\right]\,,\\
  \chi^-_{\lambda,p}(q) & = & \frac{1}{p\eta(q)}\left[
                            \lambda\Theta_{\lambda,p}(q)
                          - (\partial\Theta)_{\lambda,p}(q)\right]\,,\\
  \tilde{\chi}^{}_{\lambda,p}(q) & = & \frac{1}{\eta(q)}\left[
                            2\Theta_{\lambda,p}(q)
                            - \mathrm{i}\alpha\log(q)
                            (\partial\Theta)_{\lambda,p}(q)\right]
  \,,
\eea
where $0<\lambda<p$ and 
where we made use of the definitions (\ref{etadef}) to
(\ref{dtheta}). As before, the `logarithmic' torus amplitudes
$\tilde{\chi}_{\lambda,p}$ are not uniquely determined by these
considerations since $\alpha$ is a free constant; the form given above
is convenient for constructing modular invariant partition
functions. One should note, however, that for 
logarithmic conformal field theories the complete space of states of
the full non-chiral theory is not simply the direct sum of   
tensor products of chiral representations (see for example
\cite{Gaberdiel:1998ps}). It is therefore not clear how the full
torus amplitude has to be constructed out of these generalised
characters.  

The congruence subgroup for the $c_{p,1}$ model is $\Gamma(2p)$. There
are $2p$ characters corresponding to irreducible representations, and
$(p-1)$  `logarithmic' torus amplitudes, giving rise to a $(3p-1)$
dimensional representation of the modular group. In particular, we
therefore expect that the order of the modular differential equation
is $(3p-1)$. Furthermore, we expect that the dimension of Zhu's
algebra is $6p-1$: it follows from the 
structure of the above vacuum torus amplitudes that $p$ of the
irreducible representations have a one-dimensional ground state space,
while the other $p$ irreducible representations have ground state
multiplicity two; as above one may furthermore expect that each of the
$(p-1)$ logarithmic representations probably leads to one additional
state, thus giving altogether the dimension $p + 4p + (p -1) = 6p-1$.   

While we have not managed to write down a general expression for 
the modular differential equation for all $p$, we can give support for
these conjectures by analysing the $p=3$ triplet model with
$c=-7$. The vacuum character of this theory is $\chi^+_{2,3}(q)$. 
Under the assumption that the modular differential equation
is in fact of order $3p-1=8$, we can determine it uniquely by
requiring it to be solved by this vacuum character. Explicitly we find
\bea
0 &=& \left[
  \Big(\frac{833}{53747712}E_4(q)(E_6(q))^2
      -\frac{990437}{36691771392}(E_4(q))^4\Big)
  \right.\nonumber\\ & & \left.\mbox{}
  -\frac{40091}{143327232}(E_4(q))^2E_6(q){\it cod}_{(2)}
  \right.\nonumber\\ & & \left.\mbox{}
  +\Big(\frac{115}{746496}(E_6(q))^2
       +\frac{53467}{47775744}(E_4(q))^3\Big){\it cod}_{(4)}{\it cod}_{(2)}
  \right.\nonumber\\ & & \left.\mbox{}
  -\frac{5897}{124416}E_4(q)E_6(q){\it cod}_{(6)}{\it cod}_{(4)}
                                  {\it cod}_{(2)}
  \right.\nonumber\\ & & \left.\mbox{}
  +\frac{10889}{55296}(E_4(q))^2{\it cod}_{(8)}{\it cod}_{(6)}
                                {\it cod}_{(4)}{\it cod}_{(2)}
  \right.\nonumber\\ & & \left.\mbox{}
  +\frac{157}{432}E_6(q){\it cod}_{(10)}{\it cod}_{(8)}
                        {\it cod}_{(6)}{\it cod}_{(4)}
                        {\it cod}_{(2)}
  \right.\nonumber\\ & & \left.\mbox{}
  -\frac{21}{16}E_4(q){\it cod}_{(12)}{\it cod}_{(10)}
                      {\it cod}_{(8)}{\it cod}_{(6)}
                      {\it cod}_{(4)}{\it cod}_{(2)}
  \right.\nonumber\\ & & \left.\mbox{}
  +{\it cod}_{(16)}{\it cod}_{(14)}{\it cod}_{(12)}{\it cod}_{(10)}
   {\it cod}_{(8)}{\it cod}_{(6)}{\it cod}_{(4)}{\it cod}_{(2)}
   \vphantom{\frac{1}{2}}\right]T(q)\,.
\eea
If we make the ansatz that $T(q)$ is of the form 
\be
T(q) = q^{h-\frac{c}{24}}\, \sum_{n=0}^{\infty} \, 
\sum_{k=0}^{1} c_{k,n} \tau^k \, q^m \,, 
\ee
we obtain, to lowest order the polynomial condition
\bea
  0&=& \frac{1}{2304}(1+4h)^2h^2(h-1)(12h-5)(3h+1)(4h-7)(c_{1,0}\tau
  +c_{0,0})
  \nonumber\\
  &+&\frac{1}{1152}(1+4h)h(2304h^5-5280h^4+2160h^3+870h^2-229h-35)c_{1,0}
  \ +\ {\cal O}(q) \,.\phantom{mmm}
\eea
As expected, we can read off from this expression the 
allowed conformal weights: if the character does not involve any
powers of $\tau$ ($c_{1,0}=0$), then $h$ needs to be from the set 
$h\in\{0,-1/4,1,5/12,-1/3,7/4\}$. Furthermore, we have two
`logarithmic' torus amplitudes with $h=0$ and $h=-1/4$. This then fits
nicely together with the fact that there are in fact two 
indecomposable highest weight representations with these conformal
weights \cite{GabKau96a}.

\paragraph{Acknowledgements.} We thank Nils Carqueville and Terry
Gannon for helpful discussions. The research of MF is partially
supported by  the European Union network HPRN-CT-2002-00325 (EUCLID)
and the string theory network (SPP no.~1096), Fl 259/2-2, of the 
Deutsche Forschungsgemeinschaft. The research
of MRG is partially supported by the Swiss National Science
Foundation and the Marie Curie network `Constituents, Fundamental
Forces and Symmetries of the Universe' (MRTN-CT-2004-005104).

\end{document}